\documentclass[aps,prd,preprintnumbers,superscriptaddress,nofootinbib,showpacs,twocolumn]{revtex4}%
\usepackage[dvipdfmx]{graphicx}
\usepackage{bm,latexsym,amsmath,amssymb,amsfonts,mathrsfs}
\usepackage{color}
\input{colordvi.tex}
\allowdisplaybreaks[1]
\usepackage[dvipdfmx]{hyperref}
\usepackage{url}
\hypersetup{
    colorlinks=true,
    citecolor=cyan,
}

\begin{document}

\title{General theory of cosmological perturbations in
open and closed universes from the Horndeski action}

\author{Shingo~Akama}
\email[Email: ]{s.akama"at"rikkyo.ac.jp}
\affiliation{Department of Physics, Rikkyo University, Toshima, Tokyo 171-8501, Japan
}
\author{Tsutomu~Kobayashi}
\email[Email: ]{tsutomu"at"rikkyo.ac.jp}
\affiliation{Department of Physics, Rikkyo University, Toshima, Tokyo 171-8501, Japan
}

\begin{abstract}
Our Universe is nearly spatially flat, but this does not mean that
it is exactly spatially flat. In this paper we derive general quadratic
actions for cosmological perturbations in non-flat models from the
Horndeski theory. This allows us to study how the spatial curvature
influences the behavior of cosmological perturbations in the early universe
described by some general scalar-tensor theory.
We show that a tiny spatial curvature at the onset of inflation
is unlikely to yield large (or ${\cal O}(1)$) effects on
the primordial spectra even if one modifies gravity.
We also argue that non-singular cosmological solutions
in the Horndeski theory are unstable in
spatially open cases as well as in flat cases.
\end{abstract}

\pacs{%
98.80.Cq, 
04.50.Kd  
}
\preprint{RUP-18-31}
\maketitle

\section{Introduction}

Different cosmological observations indicate that the
Universe is nearly spatially flat.
According to Planck's observations~\cite{Aghanim:2018eyx,Akrami:2018odb},
{\it Planck} TT,TE,EE+lowE+lensing gives the constraint
in terms of the curvature density parameter $\Omega_{\cal K}$ as
\begin{align}
\Omega_{\cal K}= -0.011^{+0.013}_{-0.012}\quad (95\%\;{\rm CL}),
\end{align}
and {\it Planck} TT,TE,EE+lowE+lensing+BAO gives
\begin{align}
\Omega_{\cal K}= 0.0007\pm 0.0037\quad (95\%\;{\rm CL}).
\end{align}

The observed flatness of the universe can be explained naturally
by the inflationary expansion in
the early time~\cite{Guth:1980zm,Starobinsky:1980te,Sato:1980yn}.
For this reason it is often assumed that the universe is {\em exactly} flat.
However, the fact that $\Omega_{\cal K}$ is bounded to be small
does not mean that $\Omega_{\cal K}$ is actually equal to zero.
Some works even suggest that flat models are less favored by
data~\cite{Ooba:2017npx,Ooba:2017ukj,Ooba:2017lng,Park:2017xbl,Park:2018bwy,Park:2018fxx}.
Therefore, it is interesting to consider the possibility that
there is a tiny spatial curvature ${\cal K}\neq0$ and
\begin{align}
\left.-\frac{{\cal K}}{a^2H^2}\right|_{t=t_\ast}\sim {\Omega}_{\cal K}
= {\cal O}(10^{-3}),
\end{align}
where $H:= d\ln a/d t$ is the Hubble parameter and
$t_\ast$ is the onset of inflation or
the time at which the largest observable
scales exited the horizon during inflation.
If this is indeed the case, the dynamics of inflation
and the evolution of cosmological perturbations are
expected to be affected by non-zero ${\cal K}$.

So far the effects of the spatial curvature on the early universe
have been investigated mostly in the context of
conventional inflation driven by a canonical scalar field
with a potential~\cite{Belinsky:1987gy,Bonga:2016iuf,Bonga:2016cje,Ratra:2017ezv,Ratra:1994vw,Starobinsky:1996ek,Uzan:2003nk,Ellis:2001ym}.
However, in light of recent developments in early universe models
based on scalar-tensor theories or extended theories of gravity,
it is important to study the impact of the spatial curvature
on more general models of inflation (and possible alternative scenarios)
than previously considered.
This motivates us to formulate a general theory of
cosmological perturbations in non-flat universes,
and in this paper we do this by using the Horndeski theory,
the most general second-order scalar-tensor theory
with second-order field equations~\cite{Horndeski:1974wa}.

As applications of our general theory of cosmological perturbations,
first we explore whether it is possible or not to
yield significant effects on the primordial spectra from
a tiny spatial curvature by modifying gravity.
Second, we consider non-singular cosmology as an alternative to inflation (see, e.g.,~\cite{Battefeld:2014uga} for a review)
and discuss the stability of such cosmology with
a negative spatial curvature.

This paper is organized as follows.
In the next section we present the background equations
that govern the dynamics of a FLRW universe in the presence of
the spatial curvature.
We then derive the general quadratic actions for perturbations
of a non-flat universe from the Horndeski theory in Sec.~III.
In Sec.~IV we describe the first application of
our general theory of cosmological perturbations: inflation
in non-flat universes. In Sec.~V we analyze the stability of
non-singular open cosmologies as the second application.
Section VI is devoted to conclusions. In Appendix a further
extension of the quadratic actions for cosmological perturbations
in non-flat models is given.

\section{Background equations}

Our discussion in the main body of the paper
is based on the Horndeski theory~\cite{Horndeski:1974wa,Deffayet:2011gz,Kobayashi:2011nu},
whose action is given by
\begin{align}
S=\int d^4x\sqrt{-g}\,{\cal L}_{\rm Hor},
\end{align}
with
\begin{align}
\mathcal{L}_{\rm Hor}&=G_2(\phi,X)-G_3(\phi,X)\Box\phi
+G_4(\phi,X)R
\notag \\ &\quad
+G_{4X}\left[(\Box\phi)^2-(\nabla_{\mu}\phi\nabla_{\nu}\phi)^2\right]
\notag \\ &\quad
+G_5(\phi,X)G_{\mu\nu}\nabla^{\mu}\nabla^{\nu}\phi
-\frac{G_{5X}}{6}\bigl[(\Box\phi)^3
\notag \\ &\quad
-3\Box\phi(\nabla_{\mu}\nabla_{\nu}\phi)^2+2(\nabla_{\mu}\nabla_{\nu}\phi)^3\bigr],
\end{align}
where $X:=-g^{\mu\nu}\nabla_\mu\phi\nabla_\nu\phi/2$ and
we denote $\partial G/\partial X$ by $G_{X}$.
This action leads to the most general second-order equations of motion
for the scalar field $\phi$ and the metric $g_{\mu\nu}$,
and hence is appropriate for describing general single-field cosmology.

Although a number of spatially flat cosmological models
have been explored so far based on both
canonical and non-canonical scalar-field theories,
non-flat early universe models have been studied
mostly only in the context of the canonical scalar field.
In light of this situation, we study the effects of the spatial curvature
(denoted by ${\cal K}$ in this paper) on the background
dynamics and perturbations using the Horndeski theory.

We use the metric of the FLRW universe including
the non-flat cases,
\begin{align}
ds^2=-N^2(t)dt^2+a^2(t)\gamma_{ij}dx^idx^j,\label{metricansatz}
\end{align}
where $\gamma_{ij}$ is the metric of
maximally symmetric spatial hypersurfaces.
It can be written explicitly as
$\gamma_{ij}dx^idx^j = d\chi^2+S^2_{\cal K}(\chi)d\Omega^2$
with
\begin{align}
S_{\cal K}(\chi):=
\begin{cases}
\sin(\sqrt{\mathcal{K}}\chi)/\sqrt{\mathcal{K}}
 &({\rm closed}:\;{\cal K}>0),\\
\chi  &({\rm flat}:\;{\cal K}=0) ,\\
\sinh(\sqrt{-\mathcal{K}}\chi)/\sqrt{-\mathcal{K}}
& ({\rm open}:\;{\cal K}<0) ,\\
\end{cases}
\end{align}
and $d \Omega^2 := d\theta^2+\sin^2\theta d\varphi^2$.
Hereafter we write $(\chi, \theta, \varphi)=:\Vec{x}$.

Substituting the metric to the action,
varying it with respect to $N(t)$ and $a(t)$, and then taking $N=1$,
we obtain the background equations corresponding to
the Friedmann and evolution equations in the following form,
\begin{align}
{\cal E}_0+{\cal E}_{\cal K}&=0,
\\
{\cal P}_0+{\cal P}_{\cal K}&=0,
\end{align}
where ${\cal E}_0$ and ${\cal P}_0$ are
independent of ${\cal K}$ and ${\cal E}_{\cal K}$ and ${\cal P}_{\cal K}$
are proportional to ${\cal K}$.
They are given explicitly by
\begin{align}
{\cal E}_0&=2XG_{2X}-G_2
+6X\dot{\phi}HG_{3X}-2XG_{3\phi}-6H^2G_4
\notag \\ & \quad
+24H^2X(G_{4X}+XG_{4XX})-12HX\dot{\phi}G_{4\phi{X}}
\notag \\ &\quad
-6H\dot{\phi}G_{4\phi}
+2H^3X\dot{\phi}(5G_{5X}+2XG_{5XX})
\notag \\ &\quad
-6H^2X(3G_{5\phi}+2XG_{5\phi{X}}),
\\
{\cal P}_0&=G_2
-2X(G_{3\phi}+\ddot{\phi}G_{3X})
+2(3H^2+2\dot{H})G_4
\notag \\ & \quad
-12H^2XG_{4X}-4H\dot{X}G_{4X}-8\dot{H}XG_{4X}
\notag\\ & \quad
-8HX\dot{X}G_{4XX}
+2(\ddot{\phi}+2H\dot{\phi})G_{4\phi}+4XG_{4\phi\phi}
\notag\\ & \quad
+4X(\ddot{\phi}-2H\dot{\phi})G_{4\phi{X}}
\notag \\& \quad
-2X(2H^3\dot{\phi}+2H\dot{H}\dot{\phi}+3H^2\ddot{\phi})G_{5X}\notag\\ &\quad-4H^2X^2\ddot{\phi}G_{5XX}+4HX(\dot{X}-HX)G_{5\phi{X}}
\notag\\
&\quad
+2\left[2(HX)^{{\boldsymbol \cdot}}+3H^2X\right]G_{5\phi}+4HX\dot{\phi}G_{5\phi\phi},
\end{align}
and
\begin{align}
{\cal E}_{\cal K}=-3{\cal G}_T\frac{{\cal K}}{a^2},
\quad
{\cal P}_{\cal K}={\cal F}_T\frac{{\cal K}}{a^2},
\end{align}
with time-dependent coefficients
\begin{align}
&\mathcal{F}_T=2\left[G_4-X(\ddot{\phi}G_{5X}+G_{5\phi})\right], \label{F_T}\\
&\mathcal{G}_T=2\left[G_4-2XG_{4X}-X(H\dot{\phi}G_{5X}-G_{5\phi})\right]. \label{G_T}
\end{align}
These general equations clarify how the spatial curvature comes into play
in the background dynamics.
In general, we have
\begin{align}
\frac{{\cal E}_{\cal K}}{{\cal E}_0}\sim
\frac{{\cal P}_{\cal K}}{{\cal P}_0}\sim \frac{{\cal K}}{a^2H^2}.
\end{align}
We will see that ${\cal F}_T$ and ${\cal G}_T$ also appear
in the quadratic action for cosmological perturbations.

One can also derive the scalar-field equation
by varying the action with respect to $\phi(t)$:
\begin{align}
\frac{1}{a^3}\frac{d}{dt}\left[a^3\left(
J_0+J_{\cal K}\right)\right]=P_{\phi 0}+P_{\phi{\cal K}}, \label{eqscalar}
\end{align}
where
\begin{align}
J_0&=\dot{\phi}G_{2X}+6HXG_{3X}-2\dot{\phi}G_{3\phi}\notag\\
&\quad +6H^2\dot{\phi}(G_{4X}+2XG_{4XX})-12HXG_{4\phi{X}}\notag\\
&\quad +2H^3X(3G_{5X}+2XG_{5XX})
\notag \\ &\quad
-6H^2\dot{\phi}(G_{5\phi}+XG_{5\phi{X}}),
\\
J_{\cal K}&=6\left(\dot{\phi}G_{4X}-\dot{\phi}G_{5\phi}+HXG_{5X}\right)\frac{\mathcal{K}}{a^2},
\\
P_{\phi0}&=G_{2\phi}-2X(G_{3\phi\phi}+\ddot{\phi}G_{3\phi{X}})+6(2H^2+\dot{H})G_{4\phi}\notag\\
&\quad
+6H(\dot{X}+2HX)G_{4\phi{X}}-6H^2XG_{5\phi\phi}\notag\\
&\quad
+2H^3X\dot{\phi}G_{5\phi{X}},
\\
P_{\phi{\cal K}}&=3\frac{\partial{\cal F}_T}{\partial \phi}\frac{{\cal K}}{a^2}.
\end{align}
In contrast to the case of a minimally coupled scalar field,
the spatial curvature appears in the scalar-field equation of motion
if the scalar field is non-minimally coupled to gravity
(i.e., $G_4\neq\,$const, $G_5\neq\,$const).

The above equations were derived earlier in Ref.~\cite{Nishi:2015pta}
to study the effects of the spatial curvature
in the context of the Galilean Genesis scenario.

\section{Cosmological perturbations}\label{sec:pert}

In this section, we derive for the first time the general quadratic actions for
scalar and tensor perturbations in the presence of the spatial curvature.

Choosing the unitary gauge [$\delta\phi(t,\Vec{x})=0$],
the perturbed metric can be written as
\begin{align}
ds^2=-N^2dt^2+g_{ij}\left(dx^i+N^idt\right)\left(dx^j+N^jdt\right),
\end{align}
with
\begin{align}
&N=1+\delta{n},\quad  N_i=\mathcal{D}_i\chi,
\notag \\ &
 g_{ij}=a^2e^{2\zeta}\left(\gamma_{ij}+h_{ij}+\frac{1}{2}h_{ik}h^k_j\right),
\end{align}
where Latin indices are raised and lowered with $\gamma_{ij}$,
${\cal D}_i$ is the covariant derivative compatible with $\gamma_{ij}$, and
the tensor perturbation $h_{ij}$ satisfies $h_i^i={\cal D}_i h_j^i=0$.
Substituting this metric to the action and expanding it to
second order in perturbations, we will obtain the general
quadratic actions for tensor and scalar perturbations
in open and closed universes.
Below we will do this based on the Horndeski theory,
but the results can be extended straightforwardly to
the beyond Horndeski theories. This extension is summarized in the Appendix.

\subsection{Tensor perturbations}

With some manipulation
the general quadratic action for tensor perturbations
in open and closed universes
is found to be
\begin{align}
S^{(2)}_T=\frac{1}{8}\int{dt}d^3x\sqrt{\gamma}a^3\left[\mathcal{G}_T{\dot h_{ij}}^2+\frac{\mathcal{F}_T}{a^2}h^{ij}(\mathcal{D}^2-2\mathcal{K})h_{ij}\right],
\label{tensac2}
\end{align}
where ${\cal F}_T$ and ${\cal G}_T$ were already defined in Eqs.~(\ref{F_T})
and~(\ref{G_T}), respectively,
and \begin{math}\mathcal{D}^2:=\gamma^{ij}\mathcal{D}_i\mathcal{D}_j\end{math}.
We may define the propagation speed of tensor perturbations as
$c_t^2:={\cal F}_T/{\cal G}_T$.
Recalling that ${\cal F}_T$ and ${\cal G}_T$ do not depend
explicitly on ${\cal K}$, the effects of the spatial curvature
can be seen only in the spatial derivative operator ${\cal D}^2-2{\cal K}$.
In the case of $G_4=M_{\rm Pl}^2/2$, $G_5=0$,
the action~(\ref{tensac2}) reduces to the standard result
known in general relativity (see, e.g.,~\cite{Garriga:1997wz}).

From Eq.~(\ref{tensac2}) we see that
ghost instabilities are absent if ${\cal G}_T>0$.
Gradient instabilities can be avoided
by requiring that ${\cal F}_T/{\cal G}_T>0$.
Therefore, the stability conditions in open and closed universes are given by
\begin{align}
{\cal F}_T>0,\quad {\cal G}_T>0.
\end{align}

To canonically normalize the tensor perturbations,
we follow Ref.~\cite{Kobayashi:2011nu} and introduce a new time coordinate
\begin{align}
dy_T = \frac{{\cal F}_T^{1/2}}{a{\cal G}_T^{1/2}}dt,
\end{align}
and
\begin{align}
v_{ij}(y_T,\Vec{x})=z_T h_{ij},
\quad z_T:=\frac{a}{2}({\cal F}_T{\cal G}_T)^{1/4}.
\end{align}

In spatially flat models, it is convenient to expand the tensor perturbations
in terms of the eigenfunctions $e^{i\Vec{k}\cdot\Vec{x}}$ of the
flat-space Laplacian
and the polarization tensor.
Similarly, in open and closed models we introduce the
tensor harmonics $Q_{ij}^{nlm(s)}(\Vec{x})$
satisfying
\begin{align}
&{\cal D}^2Q_{ij}^{nlm(s)}=-k^2Q_{ij}^{nlm(s)},
\\
&{\cal D}^iQ_{ij}^{nlm(s)}=\gamma^{ij}Q_{ij}^{nlm(s)}=0.
\end{align}
Here $k^2={\cal K}(n^2-3)$ $(n=3,4,5,\cdots)$,
$2\le l \le n-1 $, $-l\le m\le l$ for ${\cal K}>0$
and $k^2=|{\cal K}|(n^2+3)$ $(n\ge 0)$, $l\ge 2$,
$-l\le m\le l$ for ${\cal K}<0$.
The index $(s)$ distinguishes between even- and odd-parity harmonics.
See Refs.~\cite{Tomita:1982ew,Abbott:1986ct} for an explicit form of $Q_{ij}^{nlm(s)}$.

We now quantize $v_{ij}$ by promoting it to the operator $\hat v_{ij}$.
Using the tensor harmonics, one can expand $\hat v_{ij}$
as\footnote{The symbol of summation should be understood as
\begin{align}
\sum_{nlm}=
\begin{cases}
\displaystyle{\sum_{n=3}^\infty \sum_{l=2}^{n-1}\sum_{m=-l}^{l} }
 &({\cal K}>0),\\
\displaystyle{ \int_0^\infty dn \sum_{l=2}^{\infty}\sum_{m=-l}^{l} }
 &({\cal K}<0) .
\end{cases}
\end{align}}
\begin{align}
\hat v_{ij}(y_T,\Vec{x})&=\sum_s \sum_{nlm}
v_{nlm}^{(s)}(y_T)\hat a_{nlm}^{(s)} Q_{ij}^{nlm(s)}(\Vec{x})
\notag \\ &\qquad \qquad \quad
+v_{nlm}^*(y_T)\hat a_{nlm}^{(s)\dagger}Q_{ij}^{nlm(s)*}(\Vec{x}),
\end{align}
where $\hat a_{nlm}^{(s)}$ and $\hat a_{nlm}^{(s)\dagger}$
are the annihilation and creation operators
satisfying the commutation relations\footnote{In the case of ${\cal K}<0$,
the Kronecker delta $\delta_{nn'}$ should be understood as
the Dirac delta function $\delta(n-n')$.}
\begin{align}
[\hat a_{nlm}^{(s)},\hat a_{n'l'm'}^{(s')\dagger}]
&=\delta_{nn'}\delta_{ll'}\delta_{mm'}\delta_{ss'},
\notag
\\
{\rm others} &= 0. \label{commute}
\end{align}
The mode functions $v_{nlm}^{(s)}$ obey
\begin{align}
{v_{nlm}^{(s)}}''+\left(|{\cal K}|n^2-{\cal K}-\frac{z_T''}{z_T}\right)v_{nlm}^{(s)}=0,
\label{mode_eq_tens}
\end{align}
where a prime denotes differentiation with respect to $y_T$.
The normalization condition is
\begin{align}
v_{nlm}^{(s)}{v_{nlm}^{(s)*}}'-{v_{nlm}^{(s)}}'v_{nlm}^{(s)*}=i.
\label{Wronskiancondition}
\end{align}

Given the background solution, one can solve Eq.~(\ref{mode_eq_tens})
with an appropriate initial condition, i.e.,
an appropriate choice of the positive frequency mode.
The primordial power spectrum can then be obtained by
evaluating
\begin{align}
{\cal P}_T(n):=\sum_{s}
\frac{|{\cal K}|^{1/2} n\left(|{\cal K}|n^2-3{\cal K}\right)}{2\pi^2}
\left|\frac{v_{nlm}^{(s)}}{z_T}\right|^2.
\end{align}
Here we have followed the definition of the power spectrum
given in~\cite{Lyth:1990dh}.

\subsection{Scalar perturbations}

The quadratic action for scalar perturbations can be obtained as
\begin{align}
S^{(2)}_S&=\int{dt}d^3x\sqrt{\gamma}a^3
\left\{-3\mathcal{G}_T\dot{\zeta}^2-\frac{\mathcal{F}_T}{a^2}\zeta\mathcal{D}^2\zeta
\right.
\notag\\&\quad
 +\Sigma \delta{n}^2-2\Theta\delta{n}\frac{\mathcal{D}^2\chi}{a^2}
  +2\mathcal{G}_T\dot{\zeta}\frac{\mathcal{D}^2\chi}{a^2}+6\Theta \delta{n}\dot{\zeta}
\notag \\ &\quad
  -2\mathcal{G}_T\delta{n}\frac{\mathcal{D}^2\zeta}{a^2}-3\mathcal{F}_T\zeta^2\frac{\mathcal{K}}{a^2}
-6\mathcal{G}_T\delta{n}\zeta\frac{\mathcal{K}}{a^2}
  \notag\\&
\left.
\quad -\frac{\mathcal{G}_T}{2a^4}
\left[(\mathcal{D}^2\chi)^2-(\mathcal{D}_i\mathcal{D}^j\chi)^2\right]
\right\}, \label{2ndscalar}
\end{align}
where
\begin{align}
\Theta=\Theta_0+\Theta_{\cal K},
\quad
\Sigma=\Sigma_0+\Sigma_{\cal K},
\end{align}
with
\begin{align}
\Theta_0&:=-\dot{\phi}XG_{3X}+2HG_4-8HXG_{4X}\notag\\ &
\quad -8HX^2G_{4XX}+\dot{\phi}G_{4\phi}+2X\dot{\phi}G_{4\phi{X}}\notag\\ &
\quad -H^2\dot{\phi}(5XG_{5X}+2X^2G_{5XX})\notag\\ &
\quad +2HX(3G_{5\phi}+2XG_{5\phi{X}}),
\\
\Theta_{\cal K}&:=-\dot{\phi}XG_{5X}\frac{\mathcal{K}}{a^2},
\\
\Sigma_0&:=XG_{2X}+2X^2G_{2XX}+12H\dot{\phi}XG_{3X}\notag\\  &
\quad +6H\dot{\phi}X^2G_{3XX}-2XG_{3\phi}-2X^2G_{3\phi{X}}-6H^2G_4\notag\\ &
\quad +6\bigl[H^2(7XG_{4X}+16X^2G_{4XX}+4X^3G_{4XXX})\notag\\ &
\quad -H\dot{\phi}(G_{4\phi}+5XG_{4\phi{X}}+2X^2G_{4\phi{X}X})\bigr]\notag\\ &
\quad +2H^3\dot{\phi}\left(15XG_{5X}+13X^2G_{5XX}+2X^3G_{5XXX}\right)\notag\\ &
\quad -6H^2X(6G_{5\phi}+9XG_{5\phi{X}}+2X^2G_{5\phi{X}X}),
\\
\Sigma_{\cal K}&:=6(XG_{4X}+2X^2G_{4XX}-XG_{5\phi}-X^2G_{5\phi{X}}\notag\\ &
\quad +2H\dot{\phi}XG_{5X}+H\dot{\phi}X^2G_{5XX})\frac{\mathcal{K}}{a^2}.
\end{align}
We can also express these quantities as
\begin{align}
\Theta_{0,{\cal K}}&=-\frac{1}{6}\frac{\partial {\cal E}_{0,{\cal K}}}{\partial H},
\\
\Sigma_{0,{\cal K}}&=X\frac{\partial {\cal E}_{0,{\cal K}}}{\partial X}
+\frac{H}{2}\frac{\partial {\cal E}_{0,{\cal K}}}{\partial H}.
\end{align}
Therefore, even in the presence of the spatial curvature
the relation between the background equation and these coefficients
remains the same as in the flat models~\cite{Kobayashi:2011nu}.

In the case of general relativity with a canonical scalar field,
$G_2=X-V(\phi)$, $G_4=M_{\rm Pl}^2/2$, $G_3=G_5=0$,
the action~(\ref{2ndscalar}) reproduces Eq.~(B.4) of
Ref.~\cite{Garriga:1997wz} in the unitary gauge.

It can be seen that
$\Theta$ and $\Sigma$ depend explicitly on ${\cal K}$,
while the other coefficients ${\cal F}_T$ and ${\cal G}_T$ do not.
Similarly to the background equations, we roughly have
\begin{align}
\frac{\Theta_{\cal K}}{\Theta_0}\sim \frac{\Sigma_{\cal K}}{\Sigma_0}
\sim \frac{{\cal K}}{a^2H^2}.
\end{align}
Note that $\Theta_{\cal K}$ and $\Sigma_{\cal K}$
are non-vanishing if $\phi$ is non-minimally coupled to gravity
except for the simplest case with $G_4=G_4(\phi)$ and $G_5=0$.
Without fine-tuning, $\Theta_{\cal K}$ and $\Sigma_{\cal K}$
cannot give rise to ${\cal O}(1)$ effects.

Variation with respect to \begin{math}\delta{n}\end{math}
and \begin{math}\chi\end{math} gives the constraint equations
\begin{align}
&\Sigma\delta{n}-\Theta\frac{\mathcal{D}^2\chi}{a^2}+3\Theta\dot{\zeta}-\mathcal{G}_T\frac{\mathcal{D}^2\zeta}{a^2}-3\mathcal{G}_T\zeta\frac{\mathcal{K}}{a^2}=0,\\
&\Theta\delta{n}-\mathcal{G}_T\dot{\zeta}-\mathcal{G}_T\frac{\mathcal{K}}{a^2}\chi=0,
\end{align}
where we used \begin{math}\mathcal{D}^2\mathcal{D}_i\chi-\mathcal{D}_i\mathcal{D}^2\chi=2\mathcal{K}\mathcal{D}_i\chi\end{math}.
Substituting these equations into Eq.~(\ref{2ndscalar}), we obtain the quadratic action for the curvature perturbation
\begin{align}
S^{(2)}_{\zeta}=\int{dt}d^3x\sqrt{\gamma}a^3\left[\mathcal{G}_S\dot{\zeta}^2+\zeta\frac{\mathcal{F}_S}{a^2}(\mathcal{D}^2+3\mathcal{K})\zeta\right],\label{reducedaction2}
\end{align}
where
\begin{align}
\mathcal{G}_S&:=
\frac{\mathcal{D}^2+3\mathcal{K}}{\mathcal{D}^2-(\mathcal{G}_T\Sigma/\Theta^2)\mathcal{K}}
\mathcal{G}_T
\left(\frac{\mathcal{G}_T\Sigma}{\Theta^2}+3\right)
 \label{G_S},\\
\mathcal{F}_S&:=\frac{1}{a}\frac{d}{dt}\biggl[
\frac{\mathcal{D}^2+3\mathcal{K}}{\mathcal{D}^2-(\mathcal{G}_T\Sigma/\Theta^2)\mathcal{K}}
\biggl(\frac{a\mathcal{G}_T^2}{\Theta}\biggr)\biggr]
\notag\\
&\quad-\mathcal{F}_T+
\frac{\mathcal{D}^2+3\mathcal{K}}{\mathcal{D}^2-(\mathcal{G}_T\Sigma/\Theta^2)\mathcal{K}}
\biggl(\frac{\mathcal{G}_T^3}{\Theta^2}\frac{\mathcal{K}}{a^2}\biggl) \label{F_S}.
\end{align}

This is the general quadratic action for $\zeta$
in open and closed universes.
In contrast to the case of the tensor perturbations,
the coefficients can depend non-trivially on ${\cal K}$
through $\Theta$ and $\Sigma$.
The squared sound speed may be defined as
$c_s^2={\cal F}_S/{\cal G}_S$, which is also
dependent on ${\cal K}$ in general.

In the case of general relativity with a canonical scalar field,
the reduced action~(\ref{reducedaction2})
reproduces the standard result found in~\cite{Gumrukcuoglu:2011zh}
written in terms of the different variable $Q:=(\dot\phi/H)\zeta$,
and,
with a non-trivial transformation of the variables
(see Appendix C of~\cite{Gumrukcuoglu:2011zh}),
the result of~\cite{Gumrukcuoglu:2011zh} can be confirmed to reproduce
the result of the $p(\phi,X)$ theory~\cite{Garriga:1999vw}
when $p=X-V(\phi)$.
However, we have not been able to
reproduce the quadratic action of~\cite{Garriga:1999vw}
for general $G_2=p(\phi,X)$ directly from Eq.~(\ref{reducedaction2}).
(Note that the definition of
our $\zeta$ is different from that of
$\zeta$ used in~\cite{Garriga:1999vw}.)

 One can derive the conditions for
avoiding gradient and ghost instabilities
based on the action~(\ref{reducedaction2}).
These instabilities are dangerous particularly for short wavelength
perturbations, because the growth rates of short modes are high.
Therefore, we take ${\cal D}^2\gg {\cal K}$ and impose
\begin{align}
{\cal G}_{S,{\rm short}}&:={\cal G}_T\left(\frac{\mathcal{G}_T\Sigma}{\Theta^2}+3\right)
>0,
\\
{\cal F}_{S,{\rm short}}&:=
\frac{1}{a}\frac{d}{dt}
\left(\frac{a\mathcal{G}_T^2}{\Theta}\right)
-\mathcal{F}_T+
\frac{\mathcal{G}_T^3}{\Theta^2}\frac{\mathcal{K}}{a^2}>0.
\end{align}

Let us move on to the quantum theory.
The quadratic action for the curvature perturbation
can be written in a canonically normalized form
by introducing
\begin{align}
dy_S=\frac{{\cal F}_S^{1/2}}{a{\cal G}_S^{1/2}}dt
\end{align}
and
\begin{align}
u(y_S,\Vec{x})=z_S\zeta,\quad z_S:=\sqrt{2}a ({\cal F}_S{\cal G}_S)^{1/4}.
\end{align}

We expand $u$ in terms of the scalar harmonics $Q^{nlm}(\Vec{x})$
satisfying
\begin{align}
{\cal D}^2Q^{nlm}=-k^2Q^{nlm},
\end{align}
where the eigenvalues are given by
$k^2=|{\cal K}|n^2-{\cal K}$
with $n=3,4,5,\cdots$, $0\le l\le n-1$, $-l\le m\le l$ (${\cal K}>0$)
and $n\ge 0$, $l\ge 0$, $-l\le m\le l$ (${\cal K}<0$).
See Refs.~\cite{Tomita:1982ew,Abbott:1986ct} for an explicit form of $Q^{nlm}$.
The operator $\hat u$ can then be expanded as
\begin{align}
\hat u(y_S,\Vec{x})&=
\sum_{nlm}u_{nlm}(y_S)\hat a_{nlm}Q^{nlm}(\Vec{x})
\notag \\ &\qquad\qquad\quad
+u_{nlm}^*(y_S)\hat a_{nlm}^\dagger Q^{nlm*}(\Vec{x}),
\end{align}
where the annihilation and creation operators $\hat a_{nlm}$ and $\hat a_{nlm}^\dagger$
satisfy the standard commutation relations.
The mode functions $u_{nlm}$ are subject to the same normalization condition
as Eq.~(\ref{Wronskiancondition}).
The equation of motion for $u_{nlm}$ is given by
\begin{align}
u_{nlm}''+\left(|{\cal K}|n^2-4{\cal K}-\frac{z_S''}{z_S}\right)u_{nlm}=0,
\end{align}
where a prime stands for differentiation with respect to $y_S$.
Following the definition given in~\cite{Lyth:1990dh},
the power spectrum for $\zeta$ is computed as
\begin{align}
{\cal P}_\zeta(n)=\frac{|{\cal K}|^{1/2}n\left(|{\cal K}|n^2-{\cal K}\right)}{2\pi^2}
\left|\frac{u_{nlm}}{z_S}\right|^2.
\end{align}

\section{Application 1: Inflation}

As an application,
let us study the case where the inflationary universe
had a spatial curvature and its effects could be seen
at the beginning of inflation.

To catch the flavor, we start with the simplest prototype with
\begin{align}
G_2=-V_0<0, \quad G_4=\frac{M^2}{2},\quad G_3=G_5=0,
\end{align}
where $V_0$ and $M$ are constants,
namely, general relativity with a positive cosmological constant.
The background equations are given by
\begin{align}
V_0-3M^2H^2-\frac{3M^2{\cal K}}{a^2}&=0,
\\
-V_0+M^2(3H^2+2\dot H)+\frac{M^2{\cal K}}{a^2}&=0,
\end{align}
which are solved by
\begin{align}
a =
\begin{cases}
a_0 \cosh(ht)\quad ({\cal K}>0),\\
a_0\sinh(ht)\quad ({\cal K}<0),
\end{cases}\label{hyperbolicsoln}
\end{align}
with
\begin{align}
h=\frac{1}{M}\sqrt{\frac{V_0}{3}}, \quad a_0h=\sqrt{|{\cal K}|}.
\end{align}
This is nothing but the de Sitter spacetime in closed/open slicing.
For $ht \gg 1$ we recover exponential expansion, $a\propto e^{ht}$,
which implies that $h$ is essentially the inflationary Hubble parameter.
However, for $ht\lesssim 1$ the evolution of the
scale factor deviates from that of the usual flat case.

In~\cite{Masso:2006gv},
the power spectrum of a test scalar field has been evaluated
for the background~(\ref{hyperbolicsoln}), without taking into account
the mixing with gravity.

Below we will consider two cases within the Horndeski theory
where the scale factor is given (approximately) by the hyperbolic functions,
but with different expressions for $a_0$ and $h$, depending on the concrete model.
Note that
if $a_0h\neq \sqrt{|{\cal K}|}$, the spacetime is
something different from de Sitter in closed/open slicing
and in particular it is
no longer de Sitter for $ht\lesssim 1$.

\subsection{Potential-driven inflation}

Let us consider the ``slow-roll'' version of the above prototype.
We assume that $\phi(t)$ moves very slowly, and expand the functions
in terms of $X$ as
\begin{align}
G_{i}&=g_{i}(\phi)+h_{i}(\phi)X+\cdots,\quad g_i \gg h_iX,
\end{align}
with $g_2(\phi)=-V(\phi)<0$. Since $g_3$ and $g_5$ can be
absorbed into the redefinition of $h_2$ and $h_4$, respectively,
after integration by parts, we may set $g_3=g_5=0$ without loss of generality.
Then, the Friedmann and evolution equations reduce to
\begin{align}
  V-6g_4H^2-6g_4\frac{{\cal K}}{a^2}&\simeq 0,
\\
-V+2g_4\left(3H^2+2\dot H\right)+2g_4\frac{{\cal K}}{a^2}&\simeq 0,
\end{align}
where we ignored the terms that vanish in the $\ddot\phi,\dot\phi\to 0$ limit,
assuming that $\ddot\phi \ll H\dot\phi$.
Although in reality we should require that ${\cal K}/a^2H^2\lesssim{\cal O}(10^{-3})$
at the onset of inflation and this might be as small as or
smaller than ignored terms, we keep the curvature terms
because we want to see how in general ${\cal K}$ appears in various equations.
From these equations we obtain
\begin{align}
H^2+\frac{{\cal K}}{a^2}&\simeq \frac{V}{6g_4},\label{sr1}
\\
\dot H-\frac{{\cal K}}{a^2}&\simeq 0.\label{sr2}
\end{align}
Therefore,
by assuming the ``slow-variation'' conditions,
$\dot g_i\ll H g_i$,
the solution of the hyperbolic form~(\ref{hyperbolicsoln})
can be obtained, but now with
\begin{align}
h=\sqrt{\frac{V}{6g_4}}\simeq {\rm const},
\quad a_0h\simeq \sqrt{|{\cal K}|},
\end{align}
as an approximate solution. This is indeed (approximately) de Sitter spacetime.
Note that in a closed universe we have $H=0$ at $t=0$ (the bounce point),
which implies that the slow-variation conditions are subtle there.
However, one does not need care about this subtle situation
as long as one focuses on the realistic case where
${\cal K}/a^2H^2$ is small enough at the onset of inflation.

The functions in the quadratic actions~(\ref{tensac2}) and~(\ref{reducedaction2})
can be evaluated as follows. It is easy to see that
\begin{align}
{\cal G}_T\simeq {\cal F}_T\simeq 2g_4\simeq {\rm const.}
\end{align}
To leading order in small quantities
(we assume that $\dot g_i/Hg_i\sim h_iX/g_i\ll1$), we obtain
\begin{align}
{\cal G}_S&\simeq \left[h_2+6h_4\left(H^2+\frac{{\cal K}}{a^2}\right)\right]\frac{X}{H^2}
\notag \\ &\quad
+6 \left[h_3+h_5\left(H^2+\frac{{\cal K}}{a^2}\right)\right]\frac{\dot\phi X}{H},
\\
{\cal F}_S&\simeq \left[h_2+6h_4\left(H^2+\frac{{\cal K}}{a^2}\right)\right]\frac{X}{H^2}
\notag \\ &\quad
+4 \left[h_3+h_5\left(H^2+\frac{{\cal K}}{a^2}\right)\right]\frac{\dot\phi X}{H}.
\end{align}
One may use the background equation to replace $H^2+{\cal K}/a^2$
with $h^2\simeq\;$const.
Therefore, the quantities inside the square brackets are approximately
constant. However, $H$ itself in the denominators can be regarded as
a constant only when $H^2 \gg {\cal K}/a^2$ and we actually have

\begin{align}
\frac{\dot{\cal G}_S}{H{\cal G}_S},\frac{\dot{\cal F}_S}{H{\cal F}_S}
\sim -\frac{\dot H}{H^2}
\sim -\frac{\cal K}{a^2H^2}.
\end{align}

As already argued above, in reality we must consider the situation
where this is indeed sufficiently small.

\subsection{Kinetically-driven inflation}

Kinetically-driven inflation is realized in the shift-symmetric theories
which are invariant under $\phi\to\phi+\rm const$.
The free functions in the Lagrangian are therefore functions of only $X$.
In shift-symmetric theories, the scalar-field equation reduces to
\begin{align}
\frac{1}{a^3}\frac{\rm d}{\rm dt}\left[a^3(J_0+J_{\cal K})\right]=0
\quad\Rightarrow\quad  J_0+J_{\cal K}=\frac{{\rm const}}{a^3}.
\end{align}
Hence, $J_0+J_{\cal K}=0$
is an attractor (as long as we focus on expanding solutions,
which is the case in this paper).
Below we study non-flat inflation
in (a certain subclass of) the shift-symmetric Horndeski theory.

The Lagrangian we consider is given by
\begin{align}
G_2=G_2(X), \quad G_4=G_4(X), \quad G_3=G_5=0.
\end{align}
We now seek for the solutions of the hyperbolic form~(\ref{hyperbolicsoln})
with $X=\;$const.
Substituting this ansatz to the Friedmann and evolution
equations, we obtain
\begin{align}
&
2XG_{2X}-G_2-6|{\cal K}_0|(G_4-2XG_{4X})
\notag \\ &-6f^2(t)[
h^2(G_4-4XG_{4X}-4X^2G_{4XX})
\notag \\ &\qquad\quad
-|{\cal K}_0|(G_4-2XG_{4X})
]=0,\label{ss1}
\\
&
G_2+2|{\cal K}_0|G_4+4h^2(G_4-2XG_{4X})
\notag \\ &+2f^2(t)\left[
h^2(G_4-2XG_{4X})-|{\cal K}_0|G_4
\right]=0,\label{ss2}
\end{align}
where ${\cal K}_0:={\cal K}/a^2_0$
and $f(t)=\tanh(ht)$ $({\cal K}>0)$,
$\coth(ht)$ (${\cal K}<0$).
The shift current vanishes for the attractor,
$J_0+J_{\cal K}=0$, leading to
\begin{align}
&G_{2X}+6|{\cal K}_0|XG_{4X}
\notag \\
& +6f^2(t)\left[
h^2(G_{4X}+2XG_{4XX})-6|{\cal K}_0|XG_{4X}
\right]=0.\label{ss3}
\end{align}
Equations~(\ref{ss1})--(\ref{ss3}) have time-dependent
and time-independent pieces. Each piece vanishes if
\begin{align}
G_2+6|{\cal K}_0|G_4&=0,
\\
G_{2X}+6|{\cal K}_0|G_{4X}&=0,
\\
h^2(G_4-2XG_{4X})-|{\cal K}_0|G_4&=0,
\\
h^2(G_{4X}+2XG_{4XX})-|{\cal K}_0|G_{4X}&=0.
\end{align}
In order for these four equations to be consistent,
\begin{align}
G_2G_{4X}-G_{2X}G_4&=0,\label{eq24-1}
\\
G_4G_{4XX}+G_{4X}^2&=0,\label{eq44-1}
\end{align}
must be satisfied for the solution $X=X_0(={\rm const})$,
and then one determines $a_0$ and $h$ as
\begin{align}
h&=\sqrt{\frac{-G_2(X_0)}{6[G_4(X_0)-2X_0G_{4X}(X_0)]}},
\label{eqa1}
\\
a_0h &=\sqrt{|{\cal K}|}\sqrt{\frac{G_4(X_0)}{G_4(X_0)-2X_0G_{4X}(X_0)}}
.\label{eqh1}
\end{align}
This is not a de Sitter spacetime because $a_0h\neq \sqrt{|{\cal K}|}$.

Note that if one included $G_3(X)$ and $G_5(X)$,
then the background equations would contain terms proportional to $H$ and $H^3$.
In that case, constructing non-flat solutions
would not be so simple as above.

As an example, let us consider a simple model with
\begin{align}
G_2=- 6 \beta G_4,\quad G_4=\frac{M_{\rm Pl}^2}{2}-\frac{\alpha}{2}X^2,
\end{align}
where $\alpha$ and $\beta$ are positive constants.
Then, Eq.~(\ref{eq24-1}) is satisfied automatically.
From Eq.~(\ref{eq44-1}) one can determine the solution $X_0$ as
\begin{align}
X_0=\frac{M_{\rm Pl}}{\sqrt{3\alpha}},
\end{align}
and from Eqs.~(\ref{eqa1}) and~(\ref{eqh1}) we obtain
\begin{align}
h= \sqrt{\frac{\beta}{3}},
\quad a_0h=\sqrt{\frac{|{\cal K|}}{3}}.
\end{align}

Note, however, that our kinetically-driven inflation
with the hyperbolic scale factor
shows a problematic behavior at the level of perturbations.
Indeed, in the above example we have
\begin{align}
{\cal G}_S=\frac{8M^2}{3}\frac{{\cal D}^2+3{\cal K}}{{\cal D}^2+5{\cal K}/3},
\quad
{\cal F}_S=\frac{2M^2}{3}\frac{4{\cal K}/3}{{\cal D}^2+5{\cal K}/3},
\end{align}
and so ${\cal F}_S\simeq 0$ for large ${\cal D}^2$.
More generically, one can check that
${\cal F}_{S,{\rm short}}=0$ irrespective of the concrete model.
This is not a surprise, because
the quadratic action for scalar perturbations becomes singular
in the (flat) de Sitter limit of usual k-inflation.
To avoid this singular behavior, we suppose that
the functions in the Lagrangian depend weakly on $\phi$
and inflation
takes place slightly away from the exact hyperbolic-type expansion.
Such a situation can be analyzed following Ref.~\cite{Saitou:2017xet}.
Then, ${\cal F}_{S,{\rm short}}$ is expected to
acquire a small (``slow-roll'' order) correction.

As for tensor perturbations we have
\begin{align}
{\cal G}_T=2M_{\rm Pl}^2,\quad {\cal F}_T=\frac{2M_{\rm Pl}^2}{3}.
\end{align}

\subsection{Primordial perturbations}

Let us evaluate the primordial power spectra
under the assumptions that the background is given by
the hyperbolic form~(\ref{hyperbolicsoln})
and ${\cal F}_T$, ${\cal G}_T$, ${\cal F}_S$,
and ${\cal G}_S$ are approximately constant.
These assumptions are analogous to those often made
in usual flat inflationary cosmology: one assumes
the de Sitter background $a\propto e^{Ht}$ and
${\cal F}_S={\cal G}_S\propto -\dot H/H^2=\dot\phi^2/2H^2=\;$const
($=$ slow-roll order $\ll 1$)
to evaluate the power spectrum analytically.

The ``$y$'' coordinate used in Sec.~\ref{sec:pert}
is given by
\begin{align}
dy={\cal C}_0\frac{dt}{a},
\end{align}
where the constant ${\cal C}_0$ depends on the concrete model
of interest as well as on the type of the perturbations.
Note that ${\cal C}_0$ corresponds the propagation speed of
the perturbations under consideration.
In closed models, this gives
\begin{align}
a(y)=\frac{a_0}{\sin(-A y)},\quad A:=\frac{a_0h}{{\cal C}_0},
\end{align}
where the range of $y$ is $-\pi/(2A)<y<0$ for $0<t<\infty$.
Since $a_0 h\sim \sqrt{|{\cal K}|}$,
$1/A$ roughly gives the effective curvature radius.
In open models, we have
\begin{align}
a(y)=\frac{a_0}{\sinh(-Ay)},
\end{align}
where the range of $y$ is $-\infty<y<0$ for $0<t<\infty$.
The evolution of the mode functions depends on $z_S''/z_S$
and $z_T''/z_T$, and now these quantities can be expressed solely
in terms of the scale factor as $z_S''/z_S\simeq a''/a$
and $z_T''/z_T\simeq a''/a$.
Thus, we want to solve the equation of the form
\begin{align}
&\psi_n''+\left(|{\cal K}|n^2-B{\cal K}-\frac{a''}{a}\right)\psi_n=0,\notag
\\
&\frac{a''}{a}=
\begin{cases}
\displaystyle{A^2\left[-1+\frac{2}{\sin^2(Ay)}\right] \quad ({\cal K}>0)},\\
\displaystyle{A^2\left[1+\frac{2}{\sinh^2(Ay)}\right]\quad ({\cal K}<0)},
\end{cases}\label{geneq11}
\end{align}
where $B=1$ for tensor perturbations and $B=4$ for scalar perturbations.
Even in the case of general relativity plus a canonical scalar field,
the ${\cal K}$-dependent part of
the equation for the curvature perturbation is different from
what is obtained for a test scalar field~\cite{Masso:2006gv}.

Equation~(\ref{geneq11}) can be solved analytically.
For ${\cal K}>0$ the general solution is given by
\begin{align}
\psi_n&=C_1\left[-\frac{A}{\tan(Ay)}+i\kappa_n\right]e^{i\kappa_n y}
\notag \\ &\quad
+C_2\left[-\frac{A}{\tan(Ay)}-i\kappa_n\right]e^{-i\kappa_n y},
\end{align}
where
\begin{align}
\kappa_n:=\sqrt{{\cal K}(n^2-B)+A^2}.
\end{align}
This is a generalization of the solutions obtained
in~\cite{Lyth:1990dh,Masso:2006gv}.
We take $C_1=0$ so that it is a positive frequency solution.
In the case of the de Sitter geometry,
this corresponds to the Bunch-Davies vacuum.
(See Refs.~\cite{Hawking:1983hj,Ratra:1984yq} for a different choice of the initial state.)
Then, it follows from the normalization condition~(\ref{Wronskiancondition})
that
\begin{align}
|C_2|^2=\frac{1}{2\kappa_n(\kappa_n^2-A^2)}.
\end{align}

We will evaluate the power spectrum in the limit $y\to0$,
so the following result will be useful:
\begin{align}
\lim_{y\to 0}\frac{|\psi_n|^2}{a^2}=\frac{A^2|C_2|^2}{a_0^2}.
\end{align}
The general solution for ${\cal K}<0$ is given by
\begin{align}
\psi_n&=C_1\left[-\frac{A}{\tanh(Ay)}+i\kappa_n\right]e^{i\kappa_n y}
\notag \\ &\quad
+C_2\left[-\frac{A}{\tanh(Ay)}-i\kappa_n\right]e^{-i\kappa_n y},
\end{align}
where
\begin{align}
\kappa_n:=\sqrt{-{\cal K}(n^2+B)-A^2}.
\end{align}
We take $C_1=0$ in order for this to be a positive frequency solution.
The normalization condition~(\ref{Wronskiancondition}) yields
\begin{align}
|C_2|^2=\frac{1}{2\kappa_n(\kappa_n^2+A^2)},
\end{align}
and also in this case we have
\begin{align}
\lim_{y\to 0}\frac{|\psi_n|^2}{a^2}=\frac{A^2|C_2|^2}{a_0^2}.
\end{align}

Using the above general formulas to evaluate the primordial spectrum
of tensor perturbations, we obtain
\begin{align}
{\cal P}_T&=\frac{2}{\pi^2}\frac{{\cal G}_T^{1/2}}{{\cal F}_T^{3/2}}h^2f_T({\cal K},n)
\notag \\ &
=\frac{2}{\pi^2}\frac{{\cal G}_T^{1/2}}{{\cal F}_T^{3/2}}
\left(H^2+\frac{a_0^2h^2}{a^2}\right)
f_T({\cal K},n),
\end{align}
where
\begin{align}
f_T({\cal K},n):=\frac{n(n^2-3\sigma_{\cal K})}
{(n^2-\sigma_{\cal K}+A^2/{\cal K})^{1/2}(n^2-\sigma_{\cal K})},
\end{align}
with $A^2=a_0^2h^2{\cal G}_T/{\cal F}_T$
and $\sigma_{\cal K}:={\rm sgn}({\cal K})=\pm 1$.
Similarly, for scalar perturbations we have
\begin{align}
{\cal P}_\zeta&=\frac{1}{8\pi^2}\frac{{\cal G}_S^{1/2}}{{\cal F}_S^{3/2}}h^2
f_S({\cal K},n)
\notag \\ &=
\frac{1}{8\pi^2}\frac{{\cal G}_S^{1/2}}{{\cal F}_S^{3/2}}
\left(H^2+\frac{a_0^2h^2}{a^2}\right)
f_S({\cal K},n),
\end{align}
where
\begin{align}
f_S({\cal K},n):=\frac{n(n^2-\sigma_{\cal K})}
{(n^2-4\sigma_{\cal K}+A^2/{\cal K})^{1/2}(n^2-4\sigma_{\cal K})},
\end{align}
with $A^2=a_0^2h^2{\cal G}_S/{\cal F}_S$.

Since the largest observable scales correspond to
$n\sim 1/\sqrt{|\Omega_{\cal K}|}$,
we generically expect that
\begin{align}
f_T, f_S = 1+{\cal O}(|\Omega_{\cal K}|),
\end{align}
but more quantitatively the correction depends on $A^2/{\cal K}$.
Let us look at potential-driven inflation as an example.
For tensor perturbations
we have $A^2/{\cal K}=\sigma_{\cal K}$,
and for scalar perturbations we have
$A^2/{\cal K}=\sigma_{\cal K}$ or $A^2/{\cal K}=(3/2)\sigma_{\cal K}$,
depending on which of $h_i$ is dominant.
Also for tensor perturbations in kinetically driven inflation
we have $A^2/{\cal K}=\sigma_{\cal K}$.
In any case, we do not find a large enhancement
factor for the ${\cal O}(|\Omega_{\cal K}|)$
corrections.
We have thus clarified how ${\cal O}(|\Omega_{\cal K}|)$
corrections enter the expressions for primordial power spectra
in the present analytic toy example.

The analytic results obtained in this subsection rely on
the assumptions that the background is given by the exact hyperbolic form
and ${\cal F}_T$, ${\cal G}_T$, ${\cal F}_S$, and ${\cal G}_S$
are constant.
More precise evaluation of the power spectra
for a given model with non-vanishing ${\cal K}$
requires numerical calculations.
However, we have already derived all the necessary basic equations
and hence performing numerical calculations is straightforward.

\section{Application 2: Stability of non-singular universes}

Our quadratic actions for cosmological perturbations
have been derived without assuming any specific background dynamics
such as inflation. Therefore, one can use the quadratic actions
for studying alternative scenarios as well.
In this section, let us consider the stability of non-singular universes.

It has been proven that
non-singular cosmological solutions in the Horndeski theory
are generally plagued with gradient instabilities
if we do not admit some pathology for tensor
perturbations~\cite{Libanov:2016kfc,Kobayashi:2016xpl,Creminelli:2016zwa}.
However, the proof assumes spatially flat models.
While it is clear that stable non-singular cosmology is
allowed in the case of closed universes,\footnote{The simplest example
is the de Sitter solution in closed slicing in general relativity
plus a cosmological constant. This implies that
instabilities of closed models are not generic and
one can avoid them by elaborating a model, suggesting that
the instability of
the closed models of non-singular cosmology in~\cite{Sberna:2017xqv}
can in principle be removed.}
in open universes it is not obvious whether or not non-singular cosmology
is possible in the Horndeski theory.
However, by using the stability conditions derived in Sec.~III
one can extend the previous no-go argument
to open models.

For open models of non-singular universes ($a\geq{\rm const}>0$, ${\cal K}<0$),
the stability conditions
$\mathcal{F}_{S,{\rm short}}>0$, ${\cal F}_T>0$, and
${\cal G}_T>0$ lead to
\begin{align}
\frac{d\xi}{dt}>a\mathcal{F}_T-\frac{\mathcal{G}_T^3}{\Theta^2}\frac{\mathcal{K}}{a^2}>0, \label{key}
\end{align}
where
\begin{align}
\xi:=\frac{a\mathcal{G}_T^2}{\Theta}.
\end{align}
From Eq.(\ref{key}) it can be seen that $\xi$ a monotonically increasing
function of $t$. Thus, the proof in~\cite{Kobayashi:2016xpl}
can be extended straightforwardly.
Note here that ${\cal K}<0$ is the essential assumption.

We integrate Eq.~(\ref{key}) from $t_i$ to $t_f$ and have
\begin{align}
\xi(t_f)-\xi(t_i)>\int^{t_f}_{t_i}{a\mathcal{F}_T} dt'+\int^{t_f}_{t_i}\frac{\mathcal{G}_T^3}{\Theta^2}\frac{(-\mathcal{K})}{a^2}dt'.
\label{cond}
\end{align}
Note that both integrands in Eq.~(\ref{cond}) are always positive.
We may suppose that $\Theta$ is finite because it
contains $H, \phi$ and $\dot\phi$. Then, $\xi$ never crosses zero,
and by taking $t_i\to-\infty$ or $t_f\to+\infty$ we see that
$\xi(\infty)-\xi(t_i)<\infty$ or $\xi(t_f)-\xi(-\infty)<\infty$ is required.
This means that both of the integrals in the right hand side of Eq.~(\ref{cond})
must be convergent for $t_i\to-\infty$ or $t_f\to +\infty$.
However, this then requires that ${\cal F}_T$ must decay sufficiently rapidly
as $t\to\infty$ or $t\to-\infty$, implying a kind of pathology
in the tensor perturbations~\cite{Kobayashi:2016xpl}
(this is directly related to geodesic incompleteness for
gravitons~\cite{Creminelli:2016zwa}).
The proof can also be applied to the case where
$\Theta$ crosses zero at some moment
and hence $\xi$ has a discontinuity there.\footnote{Note that there is
some debate about zero-crossing of $\Theta$~\cite{Battarra:2014tga,Quintin:2015rta,%
Creminelli:2016zwa,Ijjas:2017pei,Dobre:2017pnt,Mironov:2018oec,Banerjee:2018svi}.
Ijjas claimed that the unitary gauge,
which is used also in this paper, is ill-defined at $\Theta=0$~\cite{Ijjas:2017pei},
while Mironov, Rubakov, and Volkova argued that
the unitary gauge is well-defined there and the no-go theorem in the Horndeski theory
still holds even if zero-crossing of $\Theta$ occurs~\cite{Mironov:2018oec},
as is also argued in~\cite{Akama:2017jsa}.}
In this case the integrals must be convergent both for $t_i\to-\infty$
and $t_f\to +\infty$.

Thus, we have proven that all open models of non-singular universes
in the Horndeski theory are unstable
if one requires geodesic completeness for gravitons.

\section{Summary and Conclusions}

In this paper, we have formulated the non-flat inflationary dynamics
and cosmological perturbations on a non-flat background
in the Horndeski theory. We have obtained the general quadratic actions
for tensor and scalar perturbations
(see the Appendix for further generalization)
and clarified their
curvature dependence. Within the simple models we have investigated,
the corrections to the power spectra receive from the spatial curvature
are of order $\Omega_{\cal K}$ and cannot be enhanced by
modifying gravity.

Using our general quadratic actions for cosmological perturbations,
we have also studied the stability of non-singular universes
with the spatial curvature. It is obvious that one can have a stable
bouncing solution in closed models. In contrast, we have generalized
the previous no-go argument for non-singular cosmologies
in the Horndeski theory to open models.

It would be interesting to extend the effective field theory of
inflation~\cite{Cheung:2007st,Motohashi:2017gqb} to non-flat models and compare the results obtained from
the two different approaches.

\acknowledgments
We thank Tomohiro Harada, Shuichiro Yokoyama, and Shin'ichi Hirano for helpful discussions.
The work of SA was supported by the JSPS Research Fellowships for Young Scientists
No.~18J22305.
The work of TK was supported by
MEXT KAKENHI Grant Nos.~JP15H05888, JP17H06359, JP16K17707, JP18H04355,
and MEXT-Supported Program for the Strategic Research Foundation at Private Universities,
2014-2018 (S1411024).



\appendix
\section{Cosmological perturbations in non-flat universes from the beyond Horndeski action}
In this Appendix,
we extend the theory of cosmological perturbations
in non-flat models
to beyond Horndeski theories~\cite{Gleyzes:2014dya,Gao:2014soa}.
For this purpose it is convenient to use
the Lagrangian expressed in terms of the ADM variables
in the unitary gauge [$\phi(t,\Vec{x})=\phi(t)$].
We thus use the Lagrangian given by
\begin{align}
\mathcal{L}&=A_2(t,N)+A_3(t,N)K+A_4(t,N)K^2+A_5(t,N)K_{ij}^2
\notag\\ &\quad
+B_4(t,N)R+A_6(t,N)K^3+A_7(t,N)KK_{ij}^2
\notag\\ &\quad
+A_8(t,N)K_{ij}^3+B_5(t,N)K_{ij}R^{ij}+B_6(t,N)KR,\label{ADMLag1}
\end{align}
where $K_{ij}$ and $R_{ij}$ are the extrinsic and intrinsic
curvature tensors on $t=\;$const hypersurfaces (on which $\phi$ is homogeneous),
respectively. The above theory contains 10 free functions of $\phi(t)$
and $X=\dot\phi^2/(2N^2)$, which are expressed as functions of
$t$ and $N$ in Eq.~(\ref{ADMLag1}).
One may further add terms constructed from $K_{ij}$, $R_{ij}$,
and $\partial_iN/N$ in such way that the theory preserves
the three-dimensional spatial covariance~\cite{Gao:2014soa}.
However,
in order to avoid unwanted complexity,
we focus on the theory~(\ref{ADMLag1})
as a possible extension of the Horndeski theory.

When the specific relations
\begin{align}
A_5=-A_4,\quad A_7=-3A_6,\quad A_8=2A_6,\quad B_6=-\frac{1}{2}B_5
\label{rel1}
\end{align}
hold among the functions,
the Lagrangian~(\ref{ADMLag1}) reduces to that of the GLPV theory~\cite{Gleyzes:2014dya}.
When the following two conditions
\begin{align}
A_4=-B_4-N\frac{\partial{B_4}}{\partial{N}},\quad A_6=\frac{N}{6}\frac{\partial{B_5}}{\partial{N}}
\end{align}
are satisfied in addition to~(\ref{rel1}), the Horndeski theory is recovered.

Starting from the metric~(\ref{metricansatz}),
one can derive
the background equations by
varying the action with respect to $N$ and $a$:
\begin{align}
  -\mathcal{E}&:=(NA_2)'+3NA_3'H+3N^2(N^{-1}a_1)'H^2\notag\\
  &\quad+3N^3(N^{-2}a_2)'H^3+\frac{6\mathcal{K}}{a^2}(NB_4)'\notag\\
  &\quad+\frac{6\mathcal{K}}{a^2}N(B_5'+3B_6')H=0,\label{beyonde1}
  \\
\mathcal{P}&:=A_2-3a_1H^2-6a_2H^3
\notag \\ &\quad
-\frac{1}{N}\frac{d}{dt}\left(A_3+2a_1H+3a_2H^2\right)\notag\\
&\quad+\frac{2\mathcal{K}}{a^2}B_4-\frac{2\mathcal{K}}{a^2}
\frac{1}{N}\frac{d}{d t}\left(B_5+3B_6\right)=0,\label{beyondp1}
\end{align}
where the prime stands for differentiation with respect to $N$ and
we defined
\begin{align}
a_1:=3A_4+A_5,\quad a_2:=9A_6+3A_7+A_8.
\end{align}
Equations~(\ref{beyonde1}) and~(\ref{beyondp1})
can be used to determine the background evolution of $a(t)$
and $N(t)$ in the presence of the spatial curvature.

Let us now move on to deriving the quadratic actions for
cosmological perturbations.
The perturbed metric is given by
\begin{align}
ds^2=-N^2dt^2+g_{ij}(N^idt+dx^i)(N^jdt+dx^j),
\end{align}
where
\begin{align}
g_{ij}&=a^2e^{2\zeta}\left(\gamma_{ij}+h_{ij}+\frac{1}{2}\gamma^{kl}h_{ik}h_{lj}+\cdots\right),\\
N&=\overline{N}(1+\delta{n}),
\quad N_i=\overline{N}\mathcal{D}_i\chi,
\end{align}
and $\overline{N}$ is the background value of the lapse function.
To keep generality, we do not take $\overline{N}=1$.
Hereafter we will simply write the background value as $N$.

We expand the action to quadratic order in perturbations.
The quadratic action for the tensor perturbations is obtained as
\begin{align}
S_T^{(2)}=\frac{1}{8}\int{dt}d^3x\sqrt{\gamma}a^3\left[\mathcal{G}_T\dot h_{ij}^2+\frac{\mathcal{F}_T}{a^2}h^{ij}(\mathcal{D}^2-2\mathcal{K})h_{ij}\right],
\end{align}
where
\begin{align}
\mathcal{G}_T&:=2(A_5+3A_7H+3A_8H),\\
\mathcal{F}_T&:=2B_4+3B_5H+\frac{1}{N}\frac{dB_5}{dt}+6B_6H.
\end{align}
Similarly to the case of the Horndeski theory,
the explicit dependence on ${\cal K}$ appears
only in the spatial derivative operator ${\cal D}^2-2{\cal K}$.

The quadratic action for scalar perturbations is given by
\begin{align}
S_S^{(2)}=\int dt d^3x \sqrt{\gamma}{\cal L}_S,
\end{align}
with
\begin{align}
\frac{\mathcal{L}_S}{Na^3}&=-3\mathcal{G}_A\frac{\dot{\zeta}^2}{N^2}-\frac{\mathcal{F}_A}{a^2}\zeta(\mathcal{D}^2+3\mathcal{K})\zeta+\Sigma\delta{n}^2-2\Theta\delta{n}\frac{\mathcal{D}^2\chi}{a^2}\notag\\
&\quad+2\mathcal{G}_A\frac{\dot{\zeta}}{N}\frac{\mathcal{D}^2\chi}{a^2}+6\Theta\delta{n}\frac{\dot{\zeta}}{N}-2\frac{\mathcal{G}_B}{a^2}\delta{n}(\mathcal{D}^2+3\mathcal{K})\zeta\notag\\
&\quad+\mathcal{G}_T\frac{\mathcal{K}}{a^2}\chi\frac{\mathcal{D}^2\chi}{a^2}+
\frac{2\mathcal{C}_F}{a^2}(\mathcal{D}^2+3\mathcal{K})\zeta\frac{\mathcal{D}^2\chi}{a^2}-\mathcal{C}_A\frac{(\mathcal{D}^2\chi)^2}{a^4}, \label{eftscalar}
\end{align}
where
\begin{align}
\Sigma&:=NA_2'+\frac{1}{2}N^2A_2''+\frac{3H}{2}N^2A_3''\notag\\
&\quad+\frac{3}{2}\left(2a_1-2Na_1'+N^2a_1''\right)H^2\notag \\&\quad +\frac{3}{2}\left(6a_2-4Na_2'+N^2a_2''\right)H^3\notag\\
&\quad+\frac{3\mathcal{K}}{a^2}\left(2NB_4'+N^2B_4''\right)
+\frac{3\mathcal{K}}{a^2}N^2\left(B_5''+3B_6''\right)H,\\
\Theta&:=\frac{1}{2}NA_3'-(a_1-Na_1')H-\frac{3}{2}(2a_2-Na_2')H^2\notag\\
&\quad+\frac{\mathcal{K}}{a^2}N(B_5'+3B_6'),\\
\mathcal{G}_A&:=-a_1-3Ha_2,\\
\mathcal{F}_A&:=2B_4-2\frac{\dot{B_5}}{N}-6\frac{\dot{B_6}}{N},\\
\mathcal{G}_B&:=2\left(B_4+NB_4'\right)+2\left(NB_5'+3NB_6'\right)H,\\
\mathcal{C}_A&:=-A_4-A_5-\left(9A_6+5A_7+3A_8\right)H,\\
\mathcal{C}_F&:=B_5+2B_6,
\end{align}
and the relation
\begin{math}
\mathcal{G}_T=\mathcal{G}_A-3\mathcal{C}_A
\end{math}
holds.

The Euler-Lagrange equations for $\delta n$ and $\chi$
give the following constraint equations:
\begin{align}
\Sigma\delta{n}-\Theta\frac{\mathcal{D}^2\chi}{a^2}
&+3\Theta\frac{\dot \zeta}{N}-\frac{\mathcal{G}_B}{a^2}
(\mathcal{D}^2+3\mathcal{K})\zeta=0,\\
-\Theta\delta{n}+2\mathcal{G}_A\frac{\dot \zeta}{N}&+\mathcal{G}_T\frac{\mathcal{K}}{a^2}\chi-
\mathcal{C}_A\frac{\mathcal{D}^2\chi}{a^2}\notag\\&\quad
+\frac{\mathcal{C}_F}{a^2}(\mathcal{D}^2+3\mathcal{K})\zeta=0.
\label{2Lagbeyond}
\end{align}
Solving the above equations for $\delta n$ and $\chi$
and substituting the results back to~(\ref{2Lagbeyond}),
we get the reduced Lagrangian
for the curvature perturbation,
\begin{align}
\frac{\mathcal{L}_S}{Na^3}=\mathcal{G}_S\frac{{\dot \zeta}^2}{N^2}+\zeta\mathcal{F}_S\frac{(\mathcal{D}^2+3\mathcal{K})}{a^2}\zeta-\zeta\mathcal{H}_S\frac{(\mathcal{D}^2+3\mathcal{K})^2}{a^4}\zeta,
\end{align}
where
\begin{align}
\mathcal{G}_S&:=\hat {\cal O}_1
\left(3+\frac{\mathcal{G}_T\Sigma}{\Theta^2+\mathcal{C}_A\Sigma}\right)\mathcal{G}_T,
\\
\mathcal{F}_S&:=\frac{1}{Na}\frac{d}{dt}\biggl[
\hat{\cal O}_1 \frac{a\Theta\mathcal{G}_B\mathcal{G}_T}{\Theta^2+\mathcal{C}_A\Sigma}\notag\\
&\quad\quad
-\hat {\cal O}_2 a\mathcal{C}_F
\biggl(3+\frac{\mathcal{G}_T\Sigma}{\Theta^2+\mathcal{C}_A\Sigma}\biggr)
\biggr]\notag\\
&\quad\quad-\mathcal{F}_A+
\hat{\cal O}_1
\biggl(\frac{\mathcal{G}_B^2\mathcal{G}_T}{\Theta^2+\mathcal{C}_A\Sigma}\frac{\mathcal{K}}{a^2}\biggl),\\
\mathcal{H}_S&:=\frac{\hat{\cal O}_2}
{\Theta^2+\mathcal{C}_A\Sigma}\biggl(\mathcal{G}_B^2\mathcal{C}_A
+2\mathcal{G}_B\mathcal{C}_F\Theta- \mathcal{C}_F^2\Sigma\biggr),
\end{align}
with
\begin{align}
\hat {\cal O}_1&:=
\frac{\mathcal{D}^2+3\mathcal{K}}{\mathcal{D}^2-\mathcal{G}_T
\Sigma\mathcal{K}/(\Theta^2+\mathcal{C}_A\Sigma)},
\\
\hat {\cal O}_2&:=
\frac{\mathcal{D}^2 }{\mathcal{D}^2-\mathcal{G}_T
\Sigma\mathcal{K}/(\Theta^2+\mathcal{C}_A\Sigma)}.
\end{align}



\end{document}